\documentclass[reprint,amsmath,amssymb,aps,showpacs,prb,superscriptaddress]{revtex4-1}
\usepackage[dvips]{graphicx}
\usepackage{bm}
\usepackage{dcolumn}
\begin{document}

\title{Deformed aerogels in the superfluid $^3$He}


\author{I. A. Fomin}
\email{fomin@kapitza.ras.ru}
\affiliation{P. L. Kapitza Institute for Physical Problems, Russian Academy of Science, \\
Kosygina 2,
 119334 Moscow, Russia}

\author{E. V. Surovtsev}
\affiliation{P. L. Kapitza Institute for Physical Problems, Russian Academy of Science, \\
Kosygina 2,
 119334 Moscow, Russia}
\affiliation{Moscow Institute of Physics and Technologies,\\
Institutskiy per. 9, Dolgoprudny, 141700, Moscow Region, Russia}


\date{\today}

\begin{abstract}
Deformed aerogels induce a global anisotropy in the superfluid
$^3$He and orient the orbital part of its order parameter. Here a
phenomenological theory of the orientational effect of elastic
deformations of aerogels on the  superfluid phases of $^3$He in the
spirit of conventional theory of elasticity is formulated.
Phenomenological coefficients, entering basic relations, depend on
properties of given aerogel in the non-deformed state. Examples of
originally isotropic silica aerogel and axially symmetric
"nematically ordered" aerogel are considered. Values of
phenomenological coefficients are estimated with the aid of a simple
microscopic model. An example of a nonuniform deformation of
"nematically ordered" aerogel is discussed.
\end{abstract}


\maketitle

\section{Introduction}
High porosity aerogels immersed in the superfluid  $^3$He play part
of ensemble of impurities in this unconventional (p-wave)
superfluid.\cite{parp1, halp1}  For the p-wave superfluid the order
parameter is a 3$\times$3 complex matrix $A_{\mu j}$. The  first
(Greek) index of this matrix enumerates three possible projections
of spin and the second (Latin) -- three projections of the orbital
momentum of a Cooper pair. In the bulk liquid, if the effect of the
walls of a container can be neglected, the order parameter is
degenerate with respect to rotation of its orbital part. This
degeneracy can be lifted by anisotropy introduced by aerogel. There
is always a random local anisotropy with a characteristic scale of
the order of a distance between the strands forming aerogel. For a
macroscopic orientation of the order parameter of importance is the
average, or the global anisotropy with a characteristic scale of the
order of a size of a sample. Global anisotropy  depends on
composition of aerogel and technology of its preparation. Aluminium
based "nematically ordered" aerogels have strong uniaxial
anisotropy. As prepared silica aerogels (based on SiO$_2$) usually
are close to being isotropic. Anisotropy of both aerogels can be
changed by their deformation. Even small (few percent) squeezing of
the sample of silica aerogel turned out to be sufficient for fixing
orientation of the orbital vector $\mathbf{l}$ in the A-like
phase.\cite{knmts} Deformation of the "nematically ordered" aerogels
can break their axial symmetry.\cite{Dmitriev2014} At present there
are other examples of the  use of a controlled deformation of
aerogel as a tool for orientation of the order parameter of
superfluid $^3$He.\cite{Elbs, Halperin_orient, Jelev_2014} On the
other hand the existing theories are limited to analysis of a simple
uniaxial strain of originally isotropic aerogel. They are based
either on the qualitative Model of Random Cylinders (MRC),\cite{vol,
SF2007} or on a symmetry argument applied to a restricted class of
order parameters. \cite{Sauls_phases_2012} We believe that a more
systematic approach to the problem is required.

In a present paper the orientational effect of elastic macroscopic
deformations of isotropic and anisotropic aerogels on the order
parameter of superfluid  $^3$He is considered with the aid of
conventional theory of elasticity (e.g. \cite{LL_VII}) combined with
a phenomenological description of the local anisotropy induced by
aerogel.

The paper is organized as follows: Set up and phenomenological
analysis of the problem are presented in Sec. \ref{Phenomen}. In
Sec. \ref{Examples}, the proposed scheme is applied to simple
uniform deformations of isotropic and anisotropic aerogels.
Phenomenological constants, entering general expressions are
evaluated with the aid of the microscopic model in Sec. \ref{Model}
and in Sec. \ref{Twisting}, a simple example of a nonuniform
deformation is considered.

\section{\label{Phenomen}Phenomenology}

In the vicinity of the transition temperature $T_c$ presence of
aerogel in a principal order on its concentration can be formally
described by the additional term $f_a=\eta_{jl}(\textbf{r})A_{\mu
j}A_{\mu l}^*$ in the Landau expansion of the density of
thermodynamic potential, where $\eta_{jl}(\textbf{r})$ is a real
symmetric random tensor.\cite{Fomin_eta_2002} It enters the linear
part of the Ginzburg and Landau equation
\begin{equation}
[\tau\delta_{jl}+\eta_{jl}(\mathbf{r})]A_{\mu
l}-\xi^2_s\left(\frac{\partial^2 A_{\mu j}}{\partial x_l^2}+
2\frac{\partial^2 A_{\mu l}}{\partial x_l \partial x_j}\right)=0
\end{equation}
were  $\tau=(T-T_c)/T_c$ and
$\xi_s^2=\frac{7\hbar^2v_F^2\zeta(3)}{80\pi^2T_c^2}$. Tensor
$\eta_{jl}(\textbf{r})$ varies in space on a scale of the average
distance between the strands of aerogel, its isotropic part
$\frac{1}{3}\delta_{jl}\eta_{nn}(\textbf{r})$ describes local shift
of $T_c$, while the anisotropic part
$\eta^{(a)}_{jl}(\textbf{r})=\eta_{jl}(\textbf{r})-\frac{1}{3}\delta_{jl}\eta_{nn}(\textbf{r})$
- the local anisotropy. Global anisotropy of aerogel is
characterized by the ensemble average
$\left\langle\eta_{jl}(\textbf{r})\right\rangle\equiv\kappa_{jl}$.
This average varies on a distance of the order of a size of a
sample.

Deformation of a sample in terms of the conventional theory of
elasticity \cite{LL_VII} is described by the {\em displacement
vector\/} $\textbf{u}(\textbf{r})=\textbf{r}{'}-\textbf{r}$, where
$\textbf{r}$ and  $\textbf{r}{'}$ are respectively initial and final
positions of a material point within the aerogel. If the deformation
is elastic (reversible), then $\kappa_{jl}(\textbf{r})$ is a
function of derivatives  $\partial u_m/\partial x_n$. The
derivatives can be grouped in two combinations - symmetric and
antisymmetric, having different physical meaning. The symmetric
combination
\begin{equation}
u_{jl}=\frac{1}{2}\left(\frac{\partial u_j}{\partial
x_l}+\frac{\partial u_l}{\partial x_j}\right)
\end{equation}
in a linear approximation coincides with the {\em strain tensor\/}.
For given external conditions $u_{jl}$ can be found as the solution
of corresponding problem of the theory of elasticity. Then the
ensuing change of $\kappa_{jl}$ in a principal order on $u_{jl}$ can
be written as:
\begin{equation}
\label{kappa_u} \delta\kappa^{u}_{jl}=\lambda_{jlmn}u_{mn}.
\end{equation}
The ``susceptibility'' $\lambda_{jlmn}$ has obvious symmetry
properties following from the definitions of   $\kappa_{jl}$ and
$u_{mn}$:
\begin{equation}
\lambda_{jlmn}=\lambda_{ljmn}=\lambda_{jlnm}. \label{lamda_1}
\end{equation}
Additional restrictions on $\lambda_{jlmn}$ can follow from the
symmetry properties of the non-deformed aerogel.

The antisymmetric combination of derivatives
\begin{equation}
v_{jl}=\frac{1}{2}\left(\frac{\partial u_j}{\partial
x_l}-\frac{\partial u_l}{\partial x_j}\right) \label{antisym}
\end{equation}
is connected with  the local rotation of the infinitesimal volume of
aerogel at the point $\textbf{r}$ by the angle
$\delta\vec{\varphi}=\frac{1}{2}rot\mathbf{u}$. The ensuing change
of anisotropy can be written as a commutator
$\delta\kappa^{R}_{jl}=v_{jm}\kappa^{(0)}_{ml}-\kappa^{(0)}_{jm}v_{ml}$,
where $\kappa^{0}_{jl}$ is the value of tensor $\kappa_{jl}$ before
deformation. Together with Eq. (\ref{kappa_u}) it renders correction
to $\kappa_{jl}$ of the first order on derivatives of the
displacement vector:
\begin{equation}
\kappa_{jl}=\kappa_{jl}^{(0)}+v_{jm}\kappa^{(0)}_{ml}-\kappa^{(0)}_{jm}v_{ml}+\lambda_{jlmn}u_{mn}.
\label{kappa_full}
\end{equation}
This formula serves as a basis for a more detailed analysis of
particular situations.

\section{\label{Examples} Examples}
The most simple situation arises when the non-deformed aerogel is
isotropic, i.e. $\kappa_{jl}^{(0)}\sim\delta_{jl}$. Since
$\kappa_{jl}^{(0)}$ commutes with $v_{jl}$ only true deformations
$u_{jl}$ contribute to the induced anisotropy.  The susceptibility
$\lambda_{jlmn}$ in this case is a combination of isotropic tensors:
\begin{equation}
\lambda_{jlmn}=\lambda^{(0)}\delta_{jl}\delta_{mn}+\lambda^{(1)}(\delta_{jm}\delta_{ln}+\delta_{jn}\delta_{lm}-
\frac{2}{3}\delta_{jl}\delta_{mn}),                \label{lamda}
\end{equation}
where $\lambda^{(0)}$ and $\lambda^{(1)}$ are phenomenological
coefficients, and the change of $\kappa_{jl}$ is given by
\begin{equation}
\delta\kappa_{jl}=\lambda^{(0)}\delta_{jl}u_{pp}+2\lambda^{(1)}\left(u_{jl}-\frac{1}{3}\delta_{jl}u_{pp}\right).
\label{kappa_delta}
\end{equation}
Of principal interest is the second term on the r.h.s.
($\kappa^{(a)}_{jl}$ in what follows) which breaks rotational
symmetry.

As an example consider deformation of a simple extension (or simple
compression) of a sample in the form of a rectangular parallelepiped
along one of its symmetry axis, taken as z-direction. If $\delta
h_z$ is a change of its size $h_z$ in $z$-direction  only diagonal
components of $u_{jl}$ are finite:  $u_{zz}=\delta h_z/h_z,
u_{xx}=u_{yy}=-\sigma_P u_{zz}$, where $\sigma_P$ is the Poisson
ratio. Substitution of such $u_{jl}$ in Eq. (\ref{kappa_delta})
renders for the induced anisotropy tensor
\begin{equation}
\kappa^{(a)}_{jl}=\frac{2\lambda^{(1)}}{3}(1+\sigma_P)\frac{\delta
h_z}{h_z}(3\hat{z}_j\hat{z}_l-\delta_{jl}).
\end{equation}
If $^3$He is in the Anderson-Brinkman-Morel (ABM) phase with the
order parameter $A_{\mu
j}=(\Delta/\sqrt{2})d_{\mu}(\hat{m}_j+i\hat{n}_j)$ and the usual
definitions of the mutually orthogonal unit vectors
$\hat{\mathbf{m}},\hat{\mathbf{n}},\hat{\mathbf{l}}=\hat{\mathbf{m}}\times\hat{\mathbf{n}}$
the orientation energy is
\begin{equation}
\Delta\Phi=N(0)\frac{\Delta^2}{3}\lambda^{(1)}(1+\sigma_P)\frac{\delta
h_z}{h_z}[1-3(\hat{\mathbf{l}}\cdot\hat{\mathbf{z}})^2].
\end{equation}
The induced orientation of $\hat{\mathbf{l}}$ depends on the sign of
$\lambda^{(1)}$, that is determined by the concrete structure of
aerogel. In case of a simple extension $\frac{\delta h_z}{h_z}>0$.
If $\lambda^{(1)}>0$ minimum of $\Delta\Phi$ is reached at
$(\hat{\mathbf{l}}\cdot\hat{\mathbf{z}})^2=1$, i.e.
$\hat{\mathbf{l}}$ is aligned with $\hat{\mathbf{z}}$. If
$\lambda^{(1)}<0$  $\hat{\mathbf{l}}$ must belong to a plane,
perpendicular to  $\hat{\mathbf{z}}$. E.g. for aerogels used in
experiments \cite{knmts,Dmitriev2010} the latter possibility is
favored.

If deformation of the sample is a simple shear e.g. $u_y=\alpha z$,
the non-vanishing components of the strain tensor are
$u_{yz}=u_{zy}=\alpha/2$ and corresponding anisotropy tensor is
\begin{equation}
\kappa^{(a)}_{yz}=\kappa^{(a)}_{zy}=\frac{\alpha\lambda^{(1)}}{2}[\hat{y}_j\hat{z}_l+\hat{z}_j\hat{y}_l].
\label{kappa_a1}
\end{equation}
For $\lambda^{(1)}<0$ such anisotropy orients vector
$\hat{\mathbf{l}}$ of the ABM-order parameter in $yz$ plane along
the line inclined for the angle $\frac{3\pi}{4}$ with respect to
positive direction of $z$-axis. This result is expected because pure
shear is a combination of a simple extension and a simple
compression in mutually perpendicular directions. We conclude that
the anisotropy, induced in the superfluid  $^3$He by deformation of
originally isotropic aerogel is determined by one phenomenological
constant $\lambda^{(1)}$.

In the above examples we considered the average orientation of
$\hat{\mathbf{l}}$ prescribed by the global anisotropy
$\kappa^{(a)}_{jl}$. The effect of fluctuations
$\eta_{jl}(\textbf{r})-\kappa^{(a)}_{jl}$ was neglected since it
appears in the second order on the perturbation. When
$\kappa^{(a)}_{jl}=0$ only the effect of fluctuations remains.
Because of degeneracy of the state with respect to direction of
$\hat{\mathbf{l}}$ the fluctuations disrupt the long-range order and
lead to formation of the Larkin-Imry-Ma (LIM) state.\cite{IMRY_Ma,
Volovik1996} At finite $\kappa^{(a)}_{jl}$ its ordering effect
competes with the disrupting effect of fluctuations. Estimations
show\cite{vol,Fomin_robust_2008} that two effects are comparable at
a deformation  $\frac{\delta h}{h}\sim 10^{-2}-10^{-3}$. The
anisotropy wins at $\frac{\delta h}{h}\geq 10^{-2}$. If deformation
is nonuniform regions of the ordered $\hat{\mathbf{l}}$ can coexist
with regions of disrupted order.

Aluminium based aerogels \cite{Dmitriev2012} consist of long
strands, which are approximately parallel to one direction. In what
follows this direction is specified by a unit vector $\nu_j$. These
aerogels in a non-deformed state have average axial symmetry and
induce anisotropy \cite{FS2013}
\begin{equation}
\kappa^{(a0)}_{jl}=\kappa(3\hat{\nu}_j\hat{\nu}_l-\delta_{jl})\equiv
3\kappa w_{jl}.    \label{kappa_a}
\end{equation}
Experiments \cite{Dmitriev2014} show that in a presence of such
aerogel equilibrium orientation of the A-like phase corresponds to
$\hat{\mathbf{l}}\perp\hat{\mathbf{\nu}}$, i.e. to $\kappa<0$. Such
orientation of $\hat{\mathbf{l}}$ agrees with a simple physical
argument. The shorthand notation
$w_{jl}=(\hat{\nu}_j\hat{\nu}_l-\frac{1}{3}\delta_{jl})$ introduced
in Eq. (\ref{kappa_a}) is convenient for writing down a
phenomenological expression for $\lambda_{jlmn}$. Now this tensor
has to be constructed of products of components of vector
$\hat{\nu}_j$ and of the unit tensor $\delta_{mn}$. Altogether there
are six combinations, meeting the symmetry requirements specified by
Eq. (\ref{lamda_1}): $\delta_{jl}\delta_{mn};
(\delta_{jm}\delta_{ln}+\delta_{jn}\delta_{lm});
\delta_{jl}\nu_m\nu_n;
(\delta_{ln}\nu_m\nu_j+\delta_{lm}\nu_j\nu_n+\delta_{jn}\nu_m\nu_l+\delta_{jm}\nu_l\nu_n);
\delta_{mn}\nu_j\nu_l; \nu_j\nu_l\nu_m\nu_n$. It is convenient to
regroup these combinations so that
\begin{eqnarray}
\lambda_{jlmn}=\gamma_0\delta_{jl}\delta_{mn}+\gamma_1\delta_{jl}w_{mn}\nonumber\\
+\gamma_2
w_{jl}\delta_{mn}+\gamma_3 w_{jl}w_{mn}\nonumber\\
+\gamma_4(\delta_{jm}\delta_{ln}+\delta_{jn}\delta_{lm}-\frac{2}{3}\delta_{jl}\delta_{mn}-3w_{jl}w_{mn})\nonumber\\
+\gamma_5(\delta_{jm} w_{ln}+\delta_{jn} w_{lm}\nu_n+\delta_{lm}
w_{jn}+\delta_{ln}w_{jm}\nonumber\\
-\frac{4}{3}\delta_{jl} w_{mn}-\frac{4}{3}
w_{jl}\delta_{mn}-2w_{jl}w_{mn}), \label{gamma_5}
\end{eqnarray}
where  $\gamma_0,...,\gamma_5$ are phenomenological coefficients,
their relative values depend on the initial anisotropy. The first
four terms on the r.h.s. of Eq. (\ref{gamma_5}) do not break the
initial axial symmetry for arbitrary deformation, they introduce
small corrections to the overall transition temperature and to a
strength of anisotropy $\kappa$. Orientation of the order parameter
in the plane orthogonal to the symmetry axis can be induced only by
the last two terms. For example a simple extension $\frac{\delta
h_x}{h_x}$ in a direction $\hat{\mathbf{x}}$, which is perpendicular
to $\hat{\nu}_j$ corresponds to the strain tensor
$u_{mn}=\frac{\delta
h_x}{h_x}[(1+\sigma_P)\hat{x}_m\hat{x}_n-\sigma_P\delta_{mn}]$. It
leads to the increment of $\kappa_{jl}$
\begin{eqnarray}
\delta\kappa_{jl}=\frac{\delta
h_x}{h_x}\delta_{jl}[\gamma_0(1-2\sigma_P)-\gamma_1(1+\sigma_P)]\nonumber\\
+\frac{\delta h_x}{h_x}
w_{jl}[\gamma_2(1-2\sigma_P)-\gamma_3(1+\sigma_P)]\nonumber\\
+\frac{\delta
h_x}{h_x}(1+\sigma_P)\left(\gamma_4-\frac{2}{3}\gamma_5\right)(\hat{x}_j\hat{x}_l-\hat{y}_j\hat{y}_l).
\end{eqnarray}
The sense of orientation of vector $\hat{\mathbf{l}}$ of the
ABM-order parameter in the $xy$ plane is determined by the
orientational energy
\begin{equation}
\delta\Phi=N(0)\frac{\Delta^2}{2}\left(\gamma_4-\frac{2}{3}\gamma_5\right)(1+\sigma_P)\frac{\delta
h_x}{h_x}(l_y^2-l_x^2),
\end{equation}
which depends on the sign of the combination
$\left(\gamma_4-\frac{2}{3}\gamma_5\right)$. If it is positive a
uniform squeezing along $\hat{x}$-direction orients
$\hat{\mathbf{l}}$ in $\hat{y}$-direction. The only available data
for squeezing of nematic aerogel in a direction perpendicular to its
strands \cite{Dmitriev2014} favors this sense of orientation. This
result contrasts with that for isotropic silica aerogel, where
$\hat{\mathbf{l}}$ orients itself along the direction of squeezing.

For a simple shear in the $xy$-plane ($u_x=\alpha y$)
$u_{mn}=\frac{\alpha}{2}(\hat{x}_m\hat{y}_n+\hat{x}_n\hat{y}_m)$,
$v_{mn}=\frac{\alpha}{2}(\hat{x}_m\hat{y}_n-\hat{x}_n\hat{y}_m)$
$v_{jl}$ commutes with  $\kappa^{(a0)}_{jl}$
\begin{equation}
\delta\kappa_{jl}=\left(\gamma_4-\frac{2}{3}\gamma_5\right)\alpha(\hat{y}_j\hat{x}_l+\hat{x}_j\hat{y}_l)
\end{equation}
and situation is analogous to that in the isotropic aerogel. For a
shear in the $zy$-plane ($u_y=\alpha z$) both $u_{mn}$ and $v_{mn}$
contribute to the breaking of symmetry. Substitution of
$u_{mn}=\frac{\alpha}{2}(\hat{y}_m\hat{z}_n+\hat{z}_m\hat{y}_n)$ and
$v_{mn}=\frac{\alpha}{2}(\hat{y}_m\hat{z}_n-\hat{z}_m\hat{y}_n)$ in
Eq. (\ref{kappa_full}), using Eqs. (\ref{kappa_a}) and
(\ref{gamma_5}), renders
\begin{equation}
\delta\kappa_{jl}=\left(\gamma_4+\frac{1}{3}\gamma_5+\frac{3\kappa}{2}\right)\alpha(\hat{y}_j\hat{z}_l+\hat{z}_j\hat{y}_l).
\end{equation}

\section{\label{Model} Model}
The effect of deformation of aerogel on orientation of the order
parameter of  the ABM-phase of superfluid $^3$He was qualitatively
interpreted within the  "Model of Random Cylinders" (MRC in what
follows) \cite{vol} (cf. also \cite{SF2007}). Here we apply this
model to evaluation of the coefficients introduced in the previous
section. For this purpose the MRC can be formulated as follows.
Aerogel is an ensemble of randomly distributed in space and
randomly oriented identical circular cylinders.
The cylinders induce the anisotropy field $\eta_{jl}(\mathbf{r})$
which in a principal order on $\xi_0/l$, where $\xi_0$ is the
coherence length of superfluid $^3$He and $l$ is the mean free path,
can be represented as a sum of contributions of individual
cylinders:
\begin{equation}
\eta_{jl}(\mathbf{r})=\sum_s\eta_{jl}^{(1)}(\mathbf{r}-\mathbf{r}_s).
\label{eta}
\end{equation}
Here $\eta_{jl}^{(1)}(\mathbf{r}-\mathbf{r_s})$ is the local
anisotropy induced by a single cylinder, situated at the point
$\mathbf{r}_s$ and summation in Eq. (\ref{eta}) is going over all
cylinders. Averaging of Eq. (\ref{eta}) renders:
$\langle\eta_{jl}\rangle=n\langle\eta_{jl}^{(1)}(0)\rangle$, where
$n$ is the volume density of impurities and $\eta_{jl}^{(1)}(0)$ --
Fourier transform of $\eta_{jl}^{(1)}(\mathbf{r})$ at $k=0$.
Perturbation induced by
an axially symmetric impurity according to Ref.\cite{Rainer} can be
represented as a sum of two parts:
\begin{equation}
\eta_{jl}^{(1)}(0)=\frac{\pi^2}{4}\xi_0\left[\sigma^{(i)}\delta_{jl}+\sigma^{(a)}(3\hat{a}_j\hat{a}_l-\delta_{jl})\right],
\end{equation}
where  $\hat{a}_j$, $\hat{a}_l$ are components of a unit vector in
the direction of symmetry axis of a chosen impurity,
$\xi_0=\frac{\hbar v_F}{2\pi T_c}$, and $\sigma^{(i)}$ and
$\sigma^{(a)}$ are different averages of the differential
cross-section $d\sigma/d\Omega$ of Fermi quasi-particles by the
impurity. They are defined by the equality:
\begin{eqnarray}
\sigma^{(i)}\delta_{jl}+\sigma^{(a)}(3\hat{a}_j\hat{a}_l-\delta_{jl})\nonumber\\
=3\int\frac{d^2\vartheta}{4\pi}\int
d^2\vartheta'[\hat{\nu_j}\hat{\nu_l}-\hat{\nu'_j}\hat{\nu_l}]\frac{d\sigma}{d\Omega}(\mathbf{\nu},\mathbf{\nu'}).
\end{eqnarray}
E.g. if impurities are circular cylinders with the radius $\varrho$
and the height $b\gg\varrho$, which specularly reflect
quasi-particle, then (cf. \cite{SF2007})
$\sigma^{(i)}=\frac{\pi}{4}b\varrho$ and
$\sigma^{(a)}=-\frac{\pi}{8}b\varrho$. For diffusive scattering
$\sigma^{(i)}=\left(\frac{13}{18}\pi-\frac{2}{3}\right)b\varrho$ and
$\sigma^{(a)}=-\left[\left(\frac{5}{12}\right)^2\pi-\frac{1}{3}\right]b\varrho$.
Note, that the sign of $\sigma^{(a)}$ does not depend on the type of
scattering (diffusive or specular), but depends on the shape of
impurity. For the cigar-shaped impurities $\sigma^{(a)}<0$ and
$\sigma^{(a)}>0$ for the disc-like impurities.\cite{Rainer} So, we conclude, that
\begin{equation}
\langle\eta_{jl}\rangle=n\frac{\pi^2}{4}\xi_0\left[\sigma^{(i)}\delta_{jl}+\sigma^{(a)}\langle(3\hat{a}_j\hat{a}_l-\delta_{jl})\rangle\right].
 \label{eta_ave}
\end{equation}
For isotropic distribution of directions of impurities
$\langle(3\hat{a}_j\hat{a}_l-\delta_{jl})\rangle=0$. As a result the
first order correction to $T_c$ is isotropic and no global
anisotropy is induced. The shift of $T_c$ coincides with that
obtained within the conventional theory of superconducting
alloys.\cite{AG}

Deformation changes a density of aerogel $n$ and the angular average
$\langle(3\hat{a}_j\hat{a}_l-\delta_{jl})\rangle$. Relation of this
change to a macroscopic deformation has to be specified. In line
with the MRC we assume that the cylinders are "glued" in  a
continuous media. At a deformation they change their positions and
orientations together with this media. The infinitesimal vector
$dx_j$ connecting two neighboring points of the media in accord with
the definition of the displacement vector $\textbf{u}$ changes as:
\begin{equation}
dx'_j=dx_j+\frac{\partial u_j}{\partial x_m}dx_m. \label{dx}
\end{equation}
The square of a distance is expressed via the strain tensor
\begin{equation}
dx'_jdx'_j=(\delta_{mn}+2u_{mn})dx_mdx_n. \label{dxdx}
\end{equation}

For small deformations the nonlinear term in the definition of
$u_{jl}$ can be neglected. Combination of Eqs. (\ref{dx}) and
(\ref{dxdx}) in a linear approximation over derivatives of
$\textbf{u}$ renders a rule for transformation of a product of
components of the unit vector: $e_j=\frac{dx_j}{\sqrt{dx_mdx_m}}$
\begin{equation}
e'_je'_l=e_je_l+\frac{\partial u_j}{\partial
x_m}e_me_l+\frac{\partial u_l}{\partial
x_m}e_je_m-2u_{mn}e_me_ne_je_l.
\end{equation}
Using this rule we can express average
$\langle\hat{a}'_j\hat{a}'_l\rangle$ in Eq. (\ref{eta_ave}) via the
averages $\langle\hat{a}_j\hat{a}_l\rangle$ and
$\langle\hat{a}_j\hat{a}_l\hat{a}_m\hat{a}_n\rangle$ for the
non-deformed aerogel
\begin{eqnarray}
\langle\hat{a}'_j\hat{a}'_l\rangle=\langle\hat{a}_j\hat{a}_l\rangle+u_{jm}\langle\hat{a}_m\hat{a}_l\rangle+\langle\hat{a}_j\hat{a}_m\rangle
u_{ml}\nonumber\\
+v_{jm}\langle\hat{a}_m\hat{a}_l\rangle-\langle\hat{a}_j\hat{a}_m\rangle
v_{ml}-2u_{mn}\langle a_ma_na_ja_l\rangle.
\end{eqnarray}
Change of the density of impurities is expressed via the strain
tensor as:
\begin{equation}
n'=\frac{n}{1+u_{mm}}\approx n(1-u_{mm}).
\end{equation}
Together with Eq. (\ref{eta_ave}) it renders
\begin{eqnarray}
\label{etta_change}
\kappa_{jl}-\kappa_{jl}^{(0)}=\frac{\pi^2}{4}\xi_0n[-\sigma^{(i)}\delta_{jl}u_{nn}\nonumber\\
-\sigma^{(a)}\langle 3\hat{a}_j\hat{a}_l-\delta_{jl}\rangle
u_{nn}+3\sigma^{(a)}(\langle\hat{a}'_j\hat{a}'_l\rangle-
\langle\hat{a}_j\hat{a}_l\rangle)]
\end{eqnarray}

If before the deformation aerogel was isotropic
\begin{equation}
\kappa_{jl}-\kappa_{jl}^{(0)}=\frac{\pi^2}{4}\xi_0n[-\sigma^{(i)}u_{nn}\delta_{jl}+\frac{6}{5}\sigma^{(a)}(u_{jl}-
\frac{1}{3}u_{nn}\delta_{jl})]. \label{kappa_jl}
\end{equation}
In the nomenclature of the elasticity theory a trace of $u_{jl}$
represents hydrostatic compression and the traceless tensor in the
brackets represents a pure shear. Only the pure shear part can
orient the order parameter of superfluid $^3$He. Comparison of Eqs.
(\ref{kappa_jl}) and (\ref{kappa_delta}) shows that for long
cylinders the model renders $\lambda^{(1)}<0$.

In case of uniaxially symmetric aerogel with the symmetry axis
$\nu_i$ the averages $\langle\hat{a}_j\hat{a}_l\rangle$ and
$\langle\hat{a}_j\hat{a}_l\hat{a}_m\hat{a}_n\rangle$ depend on two
parameters which can be introduced as: $q_1=3\langle
a_x^2\rangle=3\langle a_y^2\rangle$, $q_2=1-3\langle
a_x^2\rangle-\langle a_z^4\rangle+\langle a_x^4\rangle$. Then

\begin{eqnarray}
\label{aveten}
\langle a_j a_k\rangle=(q_1/3)\delta_{jk}+(1-q_1)\nu_j\nu_k,\nonumber\\
\langle a_j a_l a_m a_n\rangle=(q_1/15-q_2/5)\nonumber\\
\times\left(\delta_{mn}\delta_{jl}+\delta_{jm}\delta_{ln}+\delta_{jn}\delta_{lm}\right)\nonumber\\
+q_2\cdot(\delta_{mn}\nu_j\nu_l+\delta_{jl}\nu_m\nu_n+\delta_{jm}\nu_l\nu_n+\delta_{jn}\nu_m\nu_l\nonumber\\
+\delta_{ln}\nu_m\nu_j+
\delta_{lm}\nu_n\nu_j)\nonumber\\
+(1-q_1-7q_2)\cdot\nu_j\nu_l\nu_m\nu_n.
\end{eqnarray}
From the inequalities $0\leq a_z^4\leq a_z^2\leq1$, $0\leq a_x^4\leq
a_x^2\leq1$ we find the following restrictions on $q_1$, $q_2$:
\begin{eqnarray}
\label{qq}
0\leq q_1\leq3/2,\nonumber\\
-(2/9)q_1\leq q_2\leq\min\{5/6-(7/9)q_1,q_1/3\}.
\end{eqnarray}
The case $q_1=1,~q_2=0$ corresponds to originally isotropic aerogel
$\langle a_x^2\rangle=\langle a_y^2\rangle=\langle
a_z^2\rangle=1/3$, $\langle a_x^4\rangle=\langle
a_y^4\rangle=\langle a_z^4\rangle=1/5$. For "nematically ordered"
aerogel $\langle a_z^2\rangle=\langle a_z^4\rangle=1$, i.e.
$q_1=0,~q_2=0$.

After substitution of (\ref{aveten}) into (\ref{etta_change}) one
can find phenomenological coefficients $\gamma_i$ and $\kappa$. The
first two are ${\gamma}_0=-\frac{\pi^2}{4}\xi_0n\sigma^{(i)}$, and
${\gamma}_1=0$. All remaining coefficients are proportional to
$\widetilde{\gamma}\equiv\frac{3\pi^2}{4}\xi_0n\sigma^{(a)}$:
\begin{equation}
\label{gamma}
\begin{matrix}
{\gamma}_2=(q_1-1)\widetilde{\gamma},\\
{\gamma}_3=\frac{3}{5}(q_1+12q_2)\widetilde{\gamma},\\
{\gamma}_4=\left[\frac{1}{3}-\frac{2}{15}(q_1+7q_2)\right]\widetilde{\gamma},\\
{\gamma}_5=\frac{1}{2}(1-q_1-4q_2)\widetilde{\gamma},\\
{\kappa}=\frac{1}{3}(1-q_1)\widetilde{\gamma}.
\end{matrix}
\end{equation}

The sign of the combination $\gamma_4-2/3\gamma_5$ determines a
sense of orientation of vector $\mathbf{l}$ at squeezing of axially
symmetric aerogel in a direction perpendicular to the symmetry axis.
As it was discussed in Sec. \ref{Examples}, experimental data favor
possibility $\gamma_4-2/3\gamma_5>0$. For the cigar-shaped
impurities $\sigma^{(a)}<0$ and consequently $\widetilde{\gamma}<0$.
Then it follows from the  expressions (\ref{gamma}), that inequality
$\gamma_4-2/3\gamma_5>0$ can be satisfied if  $q_2<-q_1/2$. On the
other hand one of the inequalities (\ref{qq}) is $q_2>-2/9q_1$. So
for finite $q$ orientation of vector $\mathbf{l}$ perpendicular to
the direction of squeezing is not admitted by the model. For the
limiting case of "nematically ordered" aerogel $q_1=q_2=0$ and
$\gamma_4-2/3\gamma_5=0$. It means that the model is not sufficient
even for a qualitative interpretation of the orientational effect in
this case. Essential underlying assumption of MRC is that aerogel
consist of a rigid blocks (cylinders) which do not change their form
at a macroscopic deformation of aerogel. Deformation changes only
spacial density and orientation of the blocks. This assumption may
have limited applicability to aerogels, as has been emphasized in
Ref. \onlinecite{Sauls_phases_2012}.

\section{\label{Twisting} Twisting}
As an example of a non-uniform deformation consider a twisting of a
cylinder of radius $R$. If $z$-axis is oriented along the axis of
the cylinder the displacement vector $u_j$ at twisting has following
components\cite{LL_VII}:
\begin{equation}
u_x=-\epsilon zy, u_y=\epsilon zx, u_z=0.
\end{equation}
Here $\epsilon=\frac{d\varphi}{dz}$ and $\varphi$ is the angle of
rotation of a cross-sectional plane at the deformation. The
non-vanishing components of the corresponding strain tensor are:
\begin{equation}
u_{xz}=u_{zx}=-\frac{\epsilon}{2}y,~
u_{yz}=u_{zy}=\frac{\epsilon}{2}x. \label{u_xz}
\end{equation}
Deformations can be considered as small if the condition $\epsilon R
\ll 1$ is met. Finite components of the antisymmetric tensor $v_{jl}$
defined by Eq. (\ref{antisym}) are:
\begin{equation}
v_{xy}=-\epsilon z, v_{xz}=-\frac{\epsilon}{2}y,
v_{yz}=\frac{\epsilon}{2}x. \label{u_xy}
\end{equation}
We consider situation when the original anisotropy is uniaxial and
the axis of anisotropy is parallel to the axis of the cylinder, i.e.
in the non-deformed state $\kappa^{(a0)}_{jl}=3\kappa w_{jl}$. With this $\kappa^{(a0)}_{jl}$ and Eqs.
(\ref{kappa_full}), (\ref{u_xz}) and (\ref{u_xy}) we arrive at:
\begin{eqnarray}
\label{kappa}
\delta\kappa_{xz}=\delta\kappa_{zx}=-\frac{3}{2}\widetilde{\kappa}\epsilon
y,~
\delta\kappa_{yz}=\delta\kappa_{zy}=\frac{3}{2}\widetilde{\kappa}\epsilon
x,
\end{eqnarray}
where
$\widetilde{\kappa}=\frac{2}{3}\left(\frac{3\kappa}{2}+\gamma_4+\frac{1}{3}\gamma_5\right)$.

Since the increment $\delta\kappa_{jl}$ is nonuniform the
equilibrium texture of the order parameter will be also non-uniform.
Let us consider its asymptotic form at large distances $\rho$ from
the axis of the cylinder where the gradient energy can be neglected.
Using Eq. (\ref{kappa}) the orientational energy for the ABM-phase
at the point $(x,y,z)=(\rho\cos\varphi,\rho\sin\varphi,z)$ can be
written as:
\begin{eqnarray}
\label{ani_energ} \Delta f^{(A)}=\frac{1}{2}N(0)\Delta^2\left(3\kappa w_{jk}+\delta\kappa_{jk}\right)\left(\delta_{jk}-l_jl_k\right)=\nonumber\\
=-\alpha\{3\kappa\left[\cos^2\theta-\frac{1}{3}\right]+\frac{3}{2}\widetilde{\kappa}\epsilon\rho\sin2\theta\sin(\psi-\varphi)\},
\end{eqnarray}
where $\alpha=\frac{1}{2}N(0)\Delta^2$, $\theta$ and $\psi$ are
spherical coordinates of the orbital vector
$\mathbf{l}=\mathbf{m}\times\mathbf{n}$. Minimization of Eq.
(\ref{ani_energ}) with respect to $\theta$ and $\psi$ leads to the
following two solutions:
\begin{eqnarray}
\label{texture}
\sin(\psi_{1,2}-\varphi)=\pm1, \nonumber\\
\theta_{1,2}=\frac{\pi}{2}\pm\frac{1}{2}\arctan\left(\frac{\widetilde{\kappa}}{\kappa}\epsilon\rho\right).
\end{eqnarray}
The first solution corresponds to the upper signs, and the second to
the lower ones. Substitution of either of these solutions in Eq.
(\ref{ani_energ}) renders in a principal order on $\epsilon\rho\ll
1$ the expression for the gain of orientational energy:
\begin{equation}
\label{Delta_phi} \delta
f^{(A)}\approx-\frac{3}{4}\alpha\frac{\left(\widetilde{\kappa}\rho\epsilon\right)^2}{\kappa}.
\end{equation}
Gradient terms can be neglected  if energy $\delta f^{(A)}$ is much
bigger than the density of gradient energy $f_\nabla\sim
N(0)\Delta^2(\xi_0/\rho)^2$. Using Eq. (\ref{Delta_phi}) one can
find a distance $\rho_h$ where two energies are comparable:
\begin{equation}
\rho_{h}\sim\left(\frac{\xi_0^2\kappa}{\widetilde{\kappa}^2\epsilon^2}\right)^{1/4}\sim\left(\frac{\xi_0}{n\sigma^{(a)}\epsilon^2}\right)^{1/4}.
\end{equation}
Here we assumed $\widetilde{\kappa}\sim\kappa$ and used a model
approximation $\kappa\sim n\sigma^{(a)}\xi_0$. Solutions
(\ref{texture}) can be used only in the asymptotic region
$\rho\gg\rho_{h}$.

There is another mechanism which limits from below a region of
applicability of the quasi-uniform solution. As follows from Eq.
(\ref{u_xz}) the strain vanishes at the axis of the cylinder. In the
vicinity of the axis vector $\mathbf{l}$ is oriented perpendicular
to the $z$-direction. If the rigid blocks, forming aerogel are not
axially symmetric, the fluctuating part of tensor
$\eta_{jl}(\mathbf{r})$ disrupt the long-range order of vector
$\mathbf{l}$ in the $xy$ plane and produces the two-dimensional LIM
state.\cite{Dmitriev2014}
The disordered state extends up to a distance $\rho_{LIM}$ from the
axis of the twisted cylinder, where deformation suppresses the
LIM-effect. From Eq. (\ref{kappa}) and arguments of
\cite{vol,Fomin_robust_2008} follows that this happens at
\begin{equation}
\label{Inequality2}
\rho_{LIM}\sim\frac{1}{\epsilon}\frac{\varrho^2}{\xi_0\xi_a}\left(\frac{1}{\kappa}\right)^{1/2},
\end{equation}
where $\xi_a$ is an averaged distance between the strands. For
realistic values of parameters always  $\rho_h\ll\rho_{LIM}$. It
means that  a texture of $\mathbf{l}$ can be approximately described
by Eqs. (\ref{texture}) only at $\rho>\rho_{LIM}$. This region
exists if radius of the cylinder is sufficiently large
$R>\rho_{LIM}$, or for typical values of parameters
\begin{equation}
\epsilon
R>\frac{\varrho^2}{\xi_0\xi_a}\left(\frac{1}{\kappa}\right)^{1/2}\sim5\cdot10^{-2}.
\end{equation}

Sufficiently strong magnetic field ($H\gg 30~ G$) applied parallel
to aerogel anisotropy axis orients spin vector $\mathbf{d}$
perpendicular to the field. Due to dipole energy at distances
$\rho>\rho_{LIM}$ vector $\mathbf{d}$ follows projection of orbital
vector $\mathbf{l}$ on the $xy$ plane. Therefore, at distances
$\rho>\rho_{LIM}$ one has a  texture that is close to the tangential
disgyrations both at orbital and spin spaces. Topological charge of
this type of texture is $N=1+N_1$,\cite{VW} where $N_1$ is number of
quanta of vorticity and superfluid velocity is taken to be
$\mathbf{v}_s=N_1(\hbar/2m\rho)\boldsymbol{\hat{\varphi}}$. Spin
part of the order parameter preserves vortex structure even at
distances $\rho<\rho_{LIM}$, since fluctuations of dipole energy,
caused by LIM state at orbital space, leads to LIM-length at spin
space $L^{spin}_{LIM}$ of the order of sample size
$L^{spin}_{LIM}\sim\xi_{D}(\xi_{D}/L_{LIM}^{orb})^3\sim\xi_{D}\cdot10^3\sim10$
mm, where $\xi_D$ is the dipole length $\sim10^{-2}$mm. The core
size of the d-vortex has the order of  magnetic length $\xi_{H}$.

Each of two asymptotic regions discussed above has its own NMR
"signature". If magnetic field is oriented along $z$-axis the
transverse frequency shift $\omega_{\perp}^o$ for the A-phase in the
"ordered" state can be written in a form:
\begin{eqnarray}
2\Delta\omega_{\perp}^o(\rho)\omega_L=\Omega_A^2\left(1-2\cos^2\left[\theta_{1,2}(\rho)\right]\right),
\end{eqnarray}
where $\omega_L$ is the Larmor frequency. Correction to the
frequency shift due to the spacial dependence of the texture is
small. Using Eq. (\ref{texture}) we find that
$\cos^2\left[\theta_{1,2}(\rho)\right]\sim(\epsilon\rho)^2\ll1$.
Therefore:
\begin{eqnarray}
2\Delta\omega_{\perp}^o(\rho)\omega_L\approx\Omega_A^2.
\end{eqnarray}
The frequency shift of $2D$ LIM state at the same field orientation
in the dipole-unlocked state, i.e. when $L_{LIM}^{orb}\ll\xi_D$, is equal
to:
\begin{equation}
2\Delta\omega_{\perp}^{LIM}\omega_L=\frac{1}{2}\Omega_A^2.
\end{equation}
If two states coexist, in the transverse NMR spectrum there should
be absorption within the interval
$\frac{1}{4}\frac{\Omega_A^2}{\omega_L}<\omega<\frac{1}{2}\frac{\Omega_A^2}{\omega_L}$.

\section{Conclusion}
We conclude that reaction of superfluid phases of $^3$He in aerogel
on elastic deformations of aerogel at temperatures not too far from
$T_c$ can be described in terms of few phenomenological constants -
"susceptibilities", depending on properties of the aerogel in a
non-deformed state and of $^3$He in the normal phase. The
susceptibilities relate tensor of the bulk anisotropy, which orients
the orbital parts of the order parameters of superfluid phases of
$^3$He, to the symmetric strain tensor of deformed aerogel $u_{jl}$
and antisymmetric tensor of a local rotation $v_{jl}$. These tensors
can be found with the aid of conventional theory of elasticity. At a
given pressure of helium the same phenomenological constants can be
used e.g. for the ABM and for the distorted B-like phases. Physical
mechanism of the orientational effect of a deformed aerogel is
illustrated by a simple model (MRC). This model qualitatively
describes some of the observed properties and renders realistic
estimations of values of phenomenological constants, but it is
oversimplified,  e.g. it does not explain a sense of orientation of
vector $\mathbf{l}$ for the ABM phase in "nematically ordered"
aerogel. The considered examples show that the phenomenological part
of the present analysis can serve as a framework for quantitative
interpretation of experiments with deformed aerogels.

\begin{acknowledgments}
This work was supported in part by the Russian Foundation for Basic
Research, project \# 14-02-00054-a and grant \#~MK-6180.2014.2 of
the President of the Russian Federation.
\end{acknowledgments}

\bibliography{liter1}
\end{document}